\documentclass[fleqn,12pt]{article}
\usepackage{epsfig}
\newcommand{\be}{\begin{equation}}
\newcommand{\ee}{\end{equation}}

\newcommand{\eq}[1]{eq.~(\ref{#1})} 
\newcommand{\chii}{\chi_{{}_{\rm I}}}
\newcommand{\chir}{\chi_{{}_{\rm R}}}
\newcommand{\pbar}{\bar p}
\def\signn{\sigma_{\rm nn}}
\def\siggp{\sigma_{\gamma\rm p}}
\def\siggg{\sigma_{\gamma\gamma}}
\def\rhonn{\rho_{\rm nn}}
\def\rhogp{\rho_{\gamma\rm p}}
\def\rhogg{\rho_{\gamma\gamma}}
\def\Bnn{B_{\rm nn}}
\def\Bgp{B_{\gamma\rm p}}
\def\Bgg{B_{\gamma\gamma}}
\def\sigel{\sigma_{\rm el}}
\def\sigtot{\sigma_{\rm tot}}
\renewcommand{\Im}{{\,\rm Im}}
\renewcommand{\Re}{{\,\rm Re}}
\begin{document}    
\renewcommand\thepage{\ }
%
%
\begin{titlepage} 
%
\newcommand\reportnumber{1010} 
\newcommand\mydate{August 7, 2002} 
\newlength{\nulogo} 
\settowidth{\nulogo}{\small\sf{N.U.H.E.P. Report No. \reportnumber}}
\title{\hfill\fbox{{\parbox{\nulogo}{\small\sf{Northwestern University: \\
N.U.H.E.P. Report No. \reportnumber\\ \mydate
}}}}\vspace{1in} \\
{Factorization Theorems for High Energy nn, $\gamma p$ and $\gamma\gamma$ Scattering}}
 
\author{Martin M. Block
\thanks{Work partially supported by Department of Energy contract
DA-AC02-76-Er02289 Task D.} \\
{\small\em Department of Physics and Astronomy,} \vspace{-5pt} \\ 
{\small\em Northwestern University, Evanston, IL 60208}\\
\vspace{.05in}\\
}
\vfill
\vspace{.5in}
\date{} 
\maketitle
\begin{abstract} 
\noindent The robustness of the factorization theorem for total cross sections, $\sigma_{\rm nn}/\sigma_{\gamma p}=\sigma_{\gamma p}/\sigma_{\gamma\gamma}$, originally proved by Block and Kaidalov\cite{bk} for nn (the {\em even} portion of $pp$ and $\pbar p$ scattering), $\gamma p$ and $\gamma\gamma$ scattering,  is demonstrated. Factorization theorems for the nuclear slope parameter $B$ and $\rho$, the ratio of the real to the imaginary portion of the forward scattering amplitude, are derived under very general conditions, using analyticity and the optical theorem.
\end{abstract}  
\end{titlepage} 
%
\pagenumbering{arabic}
\renewcommand{\thepage}{-- \arabic{page}\ --}  
%
Recently, Block and Kaidalov\cite{bk} have proved three high energy factorization theorems:
\begin{equation}
 \frac{\signn(s)}{\siggp(s)}=\frac{\siggp(s)}{\siggg(s)},\label{eq:sig}
\end{equation}
 where the $\sigma$'s are the total cross sections for nucleon-nucleon, $\gamma$p and $\gamma\gamma$ scattering,
\begin{equation}
\frac{\Bnn(s)}{\Bgp(s)}=\frac{\Bgp(s)}{\Bgg(s)},\label{eq:B}
\end{equation}
 where the $B$'s are the nuclear slope parameters for elastic scattering, and 
\begin{equation} \rhonn(s)=\rhogp(s)=\rhogg(s),\label{eq:rho}
\end{equation}
 where the $\rho$'s are the ratio of the real to imaginary portions of the forward scattering amplitudes,  
with the first two  factorization theorems each having their own proportionality constant. In the above, nn (for nucleon-nucleon) denotes the {\em even} portion of the $pp$ and $\pbar p$ scattering amplitude.   Their derivation assumed a eikonal model, with the  (complex) eikonal $\chi(b,s)$ such that $a(b,s)$, the (complex) scattering amplitude in impact parameter space $b$, is given by
\begin{eqnarray}
a(b,s)&=&\frac{i}{2}\left(1-e^{i\chi(b,s)}\right)\nonumber\\
&=&\frac{i}{2}\left(1-e^{-\chii(b,s)+i\chir(b,s)}
\right).\label{eik}
\end{eqnarray}
Using the optical theorem,
the total cross section $\sigma_{\rm tot}(s)$ is given by 
\begin{equation}
\sigma_{\rm tot}(s)=4\int\,d^2\vec{b} \Im a(b,s)=2\int\,\left\{1-e^{-\chii (b,s)}\cos[\chir(b,s)]\right\}\,d^2\vec{b},
\label{sigtot}
\end{equation}
and the elastic scattering cross section 
$\sigma_{\rm el}(s)$ is given by
\begin{eqnarray}
\sigma_{\rm el}(s)&=&4\int\, d^2\vec{b}\,|a(b,s)|^2=\int\left|1-e^{-\chii(b,s)+i\chir(b,s)}\right|^2\,d^2\vec{b}.
\label{sigel}
\end{eqnarray}

The ratio of the real to the imaginary portion of the forward nuclear scattering amplitude, $\rho(s)$,
is given by  
\begin{eqnarray}
\rho(s)&=&\frac{{\rm Re}\int\,d^2\vec{b}\, a(b,s)}
{{\rm Im}\int\,d^2\vec{b}\,a(b,s)}\label{rho}
\end{eqnarray}
and the nuclear slope parameter $B(s)$ is given by
\be
B(s)=
\frac{\int\,b^2a(b,s)\,d^2\vec{b}}{2\int\,a(b,s)\,d^2\vec{b}}\,.\label{Bsimple}
\ee

They used an even (under crossing) QCD-Inspired eikonal $\chi^{\rm even}$ for nn scattering, given by the sum of three contributions, glue-glue, quark-glue and quark-quark, which are individually factorizable into a product of a cross section  $\sigma (s)$ times an impact parameter space distribution function $W(b\,;\mu)$,  {\em i.e.,}:
\begin{eqnarray}
 \chi^{\rm even}(s,b)& = &\chi_{\rm gg}(s,b)+\chi_{\rm qg}(s,b)+\chi_{\rm qq}(s,b)\nonumber\\
&=&i\left[\sigma_{\rm gg}(s)W(b\,;\mu_{\rm gg})+\sigma_{\rm qg}(s)W(b\,;\mu_{\rm qg})+\sigma_{\rm qq}(s)W(b\,;\mu_{\rm qq})\right].\label{eq:chieven}
\end{eqnarray}
The impact parameter space distribution functions used in \eq{eq:chieven} were taken to be
convolutions of two dipole form factors, {\em i.e.},
\begin{equation}
W(b\,;\mu)=\frac{\mu^2}{96\pi}(\mu b)^3K_3(\mu b),\label{W}
\end{equation}
 where $K_3(x)$ is a modified Bessel function.  
For large $s$, the {\rm even} amplitude in \eq{eq:chieven} is made analytic by the substitution $s\rightarrow se^{-i\pi/2}$ (see  the table on p. 580 of reference  \cite{bc}).  

By requiring that the ratio of elastic to total scattering be process-independent, {\em i.e.,}
\be
\left(\frac{\sigma_{\rm el}}{\sigma_{\rm tot}}\right)^{\rm nn}=
\left(\frac{\sigma_{\rm el}}{\sigma_{\rm tot}}\right)^{\gamma p}=
\left(\frac{\sigma_{\rm el}}{\sigma_{\rm tot}}\right)^{\rm \gamma\gamma}\label{eq:ratio}
\ee
at {\em all} energies, a condition that insures that each process becomes equally black disk-like as we go to high energy, they showed that the eikonal $\chi^{\gamma p}$ for $\gamma p$ scattering is obtained from the substitution into $\chi^{\rm even}$ in \eq{eq:chieven} of
\be
\sigma_i\rightarrow\kappa \ {\rm and\ }\ \mu_i^2\rightarrow \mu_i^2/\kappa \label{eq:substitutegp},
\ee
and that the eikonal $\chi^{\gamma \gamma}$ for $\gamma \gamma$ scattering is found, in turn,  by making the same substitutions into $\chi^{\gamma p}$. As a consequence, they derived the three high energy factorization theorems shown above in  \eq{eq:sig}, \eq{eq:B} and \eq{eq:rho}.

It should be emphasized that the derivations in ref. \cite{bk} assumed that the impact parameter distribution functions $W(b\,;\mu)$ had the {\em same} functional forms for all three processes nn, $\gamma p$ and $\gamma\gamma$, differing only in the size of the $\mu$'s. 

The purpose of this note is to 
\begin{itemize}
\item demonstrate the robustness of the original  total cross section theorem, by using {\em different} functional forms for the $W(b)$ in the three processes,
\item  derive the $\rho$ theorem, \eq{eq:rho}, using only analyticity,
\item derive the nuclear slope theorem, \eq{eq:B}, essentially using only the optical theorem, 
\end{itemize}  
without invoking the  eikonal formalism of ref. \cite{bk} which required that all three processes---nn, $\gamma p$ and $\gamma\gamma$---used the {\em same} $W(b\,;\mu)$ functional forms.%

We will now show robustness of the original theorem, 
\begin{equation}
 \frac{\signn(s)}{\siggp(s)}=\frac{\siggp(s)}{\siggg(s)}.\label{eq:sig2}
\end{equation}
Unlike the original proof, we will now assume that the impact parameter space distributions are all different, with 
\begin{eqnarray}
W_{\rm nn}(b;\mu)&=&\frac{\mu^2}{96\pi}(\mu b)^3K_3(\mu b)\label{eq:Wnn}\\
W_{\gamma p}(b;\mu,\nu)&=&\frac{\mu^2\nu^2}{4\pi(\mu^2-\nu^2)}\left (\frac{2\mu^2}{\mu^2-\nu^2}(K_0(\nu b)-K_0(\mu b))-(\mu b)K_1(\mu b)\right)\label{eq:Wgp}\\
W_{\gamma\gamma}(b;\nu)&=&\frac{\nu^2}{4\pi}(\nu b)K_1(\nu b)\label{eq:Wgg},
\end{eqnarray}
which are the Fourier transforms
\begin{eqnarray}
W_{\rm nn}(b;\mu)&=&\frac{1}{(2\pi)^2}\int \frac{\mu^8}{(q^2+\mu^2)^4}e^{i\vec q \cdot \vec b}d^2 \vec q,\nonumber\\
W_{\gamma p}(b;\mu,\nu)&=&\frac{1}{(2\pi)^2}\int \frac{\mu^4\nu^2}{(q^2+\mu^2)^2(q^2+\nu^2)}e^{i\vec q \cdot \vec b}d^2 \vec q,\nonumber\\
W_{\gamma \gamma}(b;\nu)&=&\frac{1}{(2\pi)^2}\int \frac{\nu^4}{(q^2+\nu^2)^2}e^{i\vec q \cdot \vec b}d^2 \vec q \label{eq:fourier}
\end{eqnarray}
of a dipole-dipole, monopole-dipole and monopole-monopole form factor convolution, respectively.  The  $K$'s are modified Bessel functions.  
All of the above impact parameter space distributions are normalized such that $\int W(b)\,d^2 \vec b=1$.

We now use a $\gamma p$ form factor that is the  convolution of the dipole form factor of a nucleon with a monopole form factor of the type used for a pion, {\em i.e.,} we assume that the matter distribution in the $\gamma$ has  a similar shape to that of the  pion.  Hence, we obtain the $\gamma \gamma$ form factor  by  convoluting a monopole form factor with a monopole form factor. All that remains for our analysis is to relate the masses $\mu$ and $\nu$. To make our results effectively independent of the {\em shapes} of the distributions, we require that the mean $b^2$ using $W_{\rm nn}(b;\mu)$ of \eq{eq:Wnn} be the same as that using $W_{\gamma \gamma}(b;\mu,\nu)$ of \eq{eq:Wgg}. Hence,
\be
<b^2>=\int b^2W(b)\, d^2 \vec b=\frac{16}{\mu^2}=\frac{8}{\nu^2}\label{eq:meansquareb}
\ee
and we will use the energy-indpendent relation $
\nu=\mu/\sqrt 2.
$

\begin{figure}[hb] 
\begin{center}
\mbox{\epsfig{file=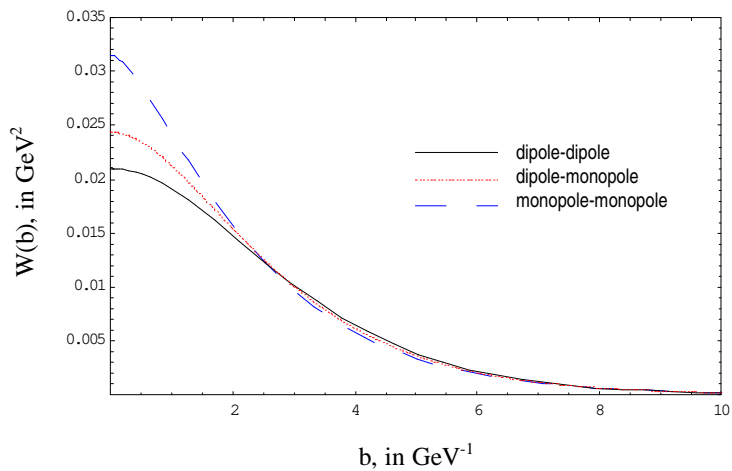,width=4.4in,%
bbllx=137pt,bblly=469pt,bburx=467pt,bbury=660pt,clip=}}
\end{center}
\protect\caption[] {\footnotesize Impact parameter space distributions $W(b)$ in GeV$^2$ {\em vs.} $b$, the impact parameter in GeV$^{-1}$. The solid curve is $W_{\rm nn}(b)$, the Fourier transform of the dipole-dipole form factor used for nn scattering,  the dotted curve is $W_{\gamma p}(b)$, the Fourier transform of the dipole-monopole form factor used for $\gamma p$ scattering and the dashed curve is $W_{\gamma \gamma}(b)$, the Fourier transform of the monopole-monopole form factor used for $\gamma \gamma$ scattering. The values of   $\mu=0.89$ GeV and  $\nu=\frac{1}{\sqrt 2}\mu=0.629$ GeV are chosen so that $<b^2>_{\rm nn}= <b^2>_{\gamma\gamma}$. 
\label{fig:eikonal}}
\end{figure}

Shown in Fig. \ref{fig:eikonal} are the  impact parameter space distributions $W(b)$ in GeV$^2$ plotted against $b$, the impact parameter in GeV$^{-1}$. The solid curve is $W_{\rm nn}(b)$, the Fourier transform of the dipole-dipole form factor used for nn scattering,  the dotted curve is $W_{\gamma p}(b)$, the Fourier transform of the dipole-monopole form factor used for $\gamma p$ scattering and the dashed curve is $W_{\gamma \gamma}(b)$, the Fourier transform of the monopole-monopole form factor used for $\gamma \gamma$ scattering, for  $\mu=0.89$ GeV, $\nu=\frac{1}{\sqrt 2}\mu = 0.629$ GeV. These choices make $<b^2>_{\rm nn}=<b^2>_{\gamma \gamma}$.

In the spirit of Block and Kaidalov\cite{bk}, we  now compare the total cross sections and the ratio of elastic to total cross sections for for the three processes, at several energies, to check for stability.  
%
%
We introduce the simpler notation  $W_{\rm nn}(b)$, $W_{\gamma p}(b)$, and $ W_{\gamma\gamma}(b)$, when \eq{eq:Wnn}, \eq{eq:Wgp} and \eq{eq:Wgg} are evaluated at $\mu=0.89$ GeV and $\nu=\frac{\mu}{\sqrt 2}= 0.629$ GeV, where $\mu=0.89$ GeV is the value for quark-quark scattering used in \cite{bhp}. For simplicity of calculation, we will assume that the eikonal is purely imaginary ($\rho =0$) and is composed of only one term, $\sigma_0\times W_{ij}(b)$, with $ij$ =nn, $\gamma p$ and $\gamma\gamma$,  respectively. Using \eq{sigtot} and \eq{sigel},  we numerically evaluate the relations
\begin{eqnarray}
\sigma_{\rm tot}&=& 2 \int\left[1-e^{-\sigma_0 W(b)}\right]\,d^2\vec b\label{eq:sigmatotal}\\
\frac{\sigma_{\rm el}}{\sigma_{\rm tot}}&=&\frac{\int\left[1-e^{-\sigma_0 W(b)}\right]^2\,d^2\vec{b}}{\sigma_{\rm tot}}\label {eq:ratio2}.
\end{eqnarray}
 We will evaluate \eq{eq:sigmatotal} and \eq{eq:ratio2} for values of $\sigma_0=160$, 110 and 67  GeV$^{-2}$, chosen so that the total cross section corresponds to the nucleon-nucleon cross section at cms energies $\approx 1500$,\  400 and 25 GeV,  respectively. The results for $\sigma_{\rm el}/\sigma_{\rm tot}$, $\sigma_{\rm tot}$ as a function of energy for the three reactions nn, $\gamma p$ and $\gamma \gamma$, as well as the energy dependence of the total cross sections $\sigma_{\rm nn}/\sigma_{\gamma p}$ and $\sigma_{\gamma p}/\sigma_{\gamma\gamma}$, are given in Table \ref{tab:energydependence}.
\begin{table}[ht]
\def\arraystretch{1.5}            
\begin{tabular}[b]{|l||l|l||l|l||l|l|}
     \cline{2-7}
      \multicolumn{1}{c|}{}
      &\multicolumn{2}{c||}{$\sqrt s=1500$ GeV}  
	&\multicolumn{2}{c||}{$\sqrt s=400$ GeV}
      &\multicolumn{2}{c|}{$\sqrt s=25$ GeV}\\
      \hline
      Reaction&$\sigma_{\rm el}/\sigma_{\rm tot} $&$\sigma_{\rm tot}$, in mb&$\sigma_{\rm el}/\sigma_{\rm tot} $&$\sigma_{\rm tot}$, in mb&$\sigma_{\rm el}/\sigma_{\rm tot} $&$\sigma_{\rm tot}$, in mb \\ \hline
     nn&0.272&73.00&0.236&57.4&0.184&40.06\\ 
     $\gamma p$&0.268&71.89&0.235&56.54&0.186&39.56\\
     $\gamma\gamma$&0.263&70.68&0.233&55.54&0.188&38.93\\ \hline\hline
\hline
      \multicolumn{1}{|c||}{$\sigma_{\rm tot}^{\rm nn}/\sigma_{\rm tot}^{\gamma p}$}
      &\multicolumn{2}{c||}{1.015}  
	&\multicolumn{2}{c||}{1.015}
      &\multicolumn{2}{c|}{1.013}\\
      \hline\multicolumn{1}{|c||}{$\sigma_{\rm tot}^{\gamma p}/\sigma_{\rm tot}^{\gamma \gamma}$}
      &\multicolumn{2}{c||}{1.017}  
	&\multicolumn{2}{c||}{1.018}
      &\multicolumn{2}{c|}{1.016}\\
      \hline
     \hline
\end{tabular}
\caption{\footnotesize The energy dependence of $\sigma_{\rm el}/\sigma_{\rm tot}$ and  $\sigma_{\rm tot}$ for the three reactions nn, $\gamma p$ and $\gamma\gamma$,  and the energy dependence of the ratios $\sigma_{\rm tot}^{\rm nn}/\sigma_{\rm tot}^{\gamma p}$ and $\sigma_{\rm tot}^{\gamma p}/\sigma_{\rm tot}^{\gamma \gamma}$. }\label{tab:energydependence}
\end{table}
For very small eikonals, the total cross sections are identical, independent of form factor shape, since all of the $W(b)$ satisfy the normalization condition $\int W(b)\, d^2\vec b=1$.

It is most striking that the total cross sections are essentially independent of the choice of form factor shape and that the ratios of elastic to total cross section are also closely  the same for all three processes, independent of energy, when we equate the $<b^2>$'s,  the mean squared $b$ values for nn and $\gamma \gamma$ processes. We further note that the ratios $\sigma_{\rm tot}^{\rm nn}/\sigma_{\rm tot}^{\gamma p}$ and $\sigma_{\rm tot}^{\gamma p}/\sigma_{\rm tot}^{\gamma \gamma}$ are {\em both  systematically} $\approx  1.5$ \% too large. As a consequence, the original cross section factorization theorem of Block and Kaidalov\cite{bk},
\begin{equation}
 \frac{\signn(s)}{\siggp(s)}=\frac{\siggp(s)}{\siggg(s)},\label{eq:sig3}
\end{equation}
is shown to be valid to $\approx 0.3$ \% over a very large range of $s$.   Thus, even without the constraint of identical $W(b)$ distributions for the three processes, the cross section factorization theorem is very robust.

We now turn our attention to the factorization theorems involving $\rho$ and $B$.      
We will find $\rho_{\gamma\gamma}(s)$ utilizing an analysis involving real analytic amplitudes, a technique first proposed by Bourrely and Fischer\cite{bourrely} and later utilized extensively by Nicolescu and Kang\cite{kang}. We follow the procedures and conventions used by Block and Cahn\cite{bc}. The variable $s$ is the square of the c.m. system energy, whereas $\nu$ is the laboratory system momentum. Using the optical theorem, in terms of the {\em even}  laboratory scattering amplitude $f_+$, where $f_+(\nu)=f_+(-\nu)$, the
total even cross section $\sigtot$ is given by
\begin{eqnarray}
\sigtot&=&\frac{4\pi}{\nu}{\rm Im}f_+(\theta_{\rm lab}=0),\label{optical}
\end{eqnarray}
where $\theta_{\rm lab}$ is the laboratory scattering angle.
We further assume that our amplitudes are real analytic functions with a simple cut structure\cite{bc}. We use an even amplitude for reactions in the high energy region, far above any cuts,  (see ref.\cite{bc}, p. 587, eq. (5.5a), with $a=0$), where the even amplitude simplifies considerably and is given, for example,  by
\begin{equation}
f_+(s)=i\frac{\nu}{4\pi}\left\{A+\beta[\ln (s/s_0) -i\pi/2]^2+cs^{\mu-1}e^{i\pi(1-\mu)/2}\right\},\label{evenamplitude_gp}
\end{equation}
where $A$, $\beta$, $c$, $s_0$ and $\mu$ are real constants. We are ignoring any real subtraction constants. In \eq{evenamplitude_gp}, we have assumed that the total cross section rises asymptotically as $\ln^2 s$. The real and imaginary parts of \eq{evenamplitude_gp} are given by
\begin{eqnarray}
{\rm Re}\frac{4\pi}{\nu}f_+(s) &=&\beta\,\pi \ln s/s_0-c\,\cos(\pi\mu/2)s^{\mu-1}\label{real}\\ 
{\rm Im}\frac{4\pi}{\nu}f_+(s) &=&A+\beta\left[\ln^2 s/s_0-\frac{\pi^2}{4}\right]+c\,\sin(\pi\mu/2)s^{\mu-1}. \label{imaginary}
\end{eqnarray}
 Using equations ({\ref{optical}), (\ref{real}) and  (\ref{imaginary}), the total cross section for high energy scattering is given by
\be
\sigtot(s)= A+\beta\left[\ln^2 s/s_0-\frac{\pi^2}{4}\right]+c\,\sin(\pi\mu/2)s^{\mu-1} , \label{sigmatot}
\ee 
and $\rho$, the ratio of the real to the imaginary portion of the forward scattering amplitude, is given by
\be
\rho(s)=\frac{\beta\,\pi\ln s/s_0-c\,\cos(\pi\mu/2)s^{\mu-1}}{A+\beta\left[\ln^2 s/s_0-\frac{\pi^2}{4}\right]+c\,\sin(\pi\mu/2)s^{\mu-1}}=\frac{\beta\,\pi\ln s/s_0-c\,\cos(\pi\mu/2)s^{\mu-1}}{\sigtot}.\label{rho2}
\ee  
Clearly, if the cross sections for the three processes, nn, $\gamma p$, and $\gamma\gamma$ scale, {\em i.e.}, $\frac{\signn(s)}{\siggp(s)}=\frac{\siggp(s)}{\siggg(s)}$, then by inspection of \eq{sigmatot}, the coefficients $A,\beta$ and $c$ for each of the three processes scale.  Thus, the scale factor cancels out in $\rho$, which is a ratio, as seen in \eq{rho2}.  Hence, all three $\rho$ values are the {\em same}, {\em i.e.}, 
\be\rhonn(s)=\rhogp(s)=\rhogg(s),\ee
 the $\rho$ factorization theorem of \eq{eq:rho}. Clearly, the argument does not depend on the specific form  of the amplitude assumed in \eq{evenamplitude_gp}, being equally valid if the cross section were to rise as $\ln s$, or even rise as a power of $s$. 

Let us now turn our attention to the factorization theorem of the nuclear slopes.  We will assume that the differential elastic scattering is adequately parameterized by 
\be
\frac{d\sigel}{dt}=\left[\frac{d\sigel}{dt} \right]_{t=0}e^{Bt}.\label{eq:dsdt0}
\ee
We can write, working in the center of mass system,   
\begin{eqnarray}
\left[\frac{d\sigel}{dt} \right]_{t=0}&=&\frac{\pi}{k^2}\left[\frac{d\sigel}{d\Omega_{\rm c.m.}} \right]_{\theta_{\rm c.m.}=0}\nonumber\\
&=&\frac{\pi}{k^2}\left|\Re f_{\rm c.m.}(0)+i\Im f_{\rm c.m.}(0)\right|^2.\label{eq:dsdt1}
\end{eqnarray}
Introducing $\rho=\Re f_{\rm c.m.}(0)/\Im f_{\rm c.m.}(0)$, we rewrite \eq{eq:dsdt1} as
\begin{eqnarray}
\left[\frac{d\sigel}{dt} \right]_{t=0}&=&\pi\left|\frac{(\rho + i)\Im f_{\rm c.m.}(0)}{k}\right|^2\nonumber\\
&=&\pi\left|\frac{(\rho+i)\sigtot}{4\pi}\right|^2\nonumber\\
&=&\frac{(1+\rho^2)\sigtot^2}{16\pi},\label{eq:dsdt2}
\end{eqnarray}
where we have used the optical theorem, $\sigtot=\frac{4\pi}{k}\Im f_{\rm c.m.}(0)$, where $k=$ the center of mass momentum,   in the next-to-last step.
Integrating  \eq{eq:dsdt0} over $t$ from $-\infty$ to $0$, we find the total elastic scattering
\be
\sigel = \frac{\sigtot^2(1+\rho^2)}{16\pi B}\label{eq:sigmael}.
\ee
We can finally rewrite \eq{eq:sigmael} in the useful form
\be
\frac{\sigel}{\sigtot} = \frac{\sigtot(1+\rho^2)}{16\pi B}\label{eq:sigeloversigtot}.
\ee
It should be pointed out that the application of \eq{eq:sigeloversigtot} to $\gamma p$  and $\gamma \gamma$ processes assumes that  the photon has turned into a hadron and is interacting hadronically.

If we apply to \eq{eq:sigeloversigtot} the fundamental condition used by Block and Kaidalov\cite{bk} that the ratio of elastic to total cross section is process-independent---after using the equality of the $\rho$ values for all three processes---we find that
\be
\frac{\signn}{\Bnn}=\frac{\siggp}{\Bgp}=\frac{\siggg}{\Bgg}.\label{eq:sigoverB}
\ee
Applying the cross section factorization theorem of \eq{eq:sig},   
 $\signn(s)/\siggp(s)=\siggp(s)/\siggg(s)$, to \eq{eq:sigoverB}, we deduce the factorization theorem of \eq{eq:B} for the nuclear slopes $B$, {\em i.e.}, 
\be
\frac{\Bnn(s)}{\Bgp(s)}=\frac{\Bgp(s)}{\Bgg(s)}.\label{eq:B2}
\ee

In conclusion, we have demonstrated the robustness of the factorization theorem for total cross sections, even when using different shapes of the impact parameter distribution functions for nn, $\gamma p$ and $\gamma \gamma$ scattering. Further, we have proved the $B$ and the $\rho$ factorization theorems of Block and Kaidalov\cite{bk} using the more general conditions of the optical theorem and analyticity. Experimental evidence for these theorems can be  found in references  \cite{bhp},\cite{rhogp} and \cite{rhogg}. 

I would like to acknowledge the hospitality of the Aspen Center for Physics during the preparation of this manuscript.


\begin{thebibliography}{99} 
     \vspace{.3in}
%
\bibitem{bk}M. M. Block and A. B. Kaidalov, e-Print Archive: {\bf hep-ph/0012365}, Phys. Rev. D {\bf 64}, 076002 (2001).

\bibitem{bourrely} C.~Bourrely and J.~Fischer, Nucl. Phys. B {\bf 61}, 513 (1973).
%
\bibitem{kang}
L.Lukaszuk and B. Nicolescu, Lett. Nuovo Cimento {\bf 8}, 405(1973) ;
K. Kang and B. Nicolescu, Phys. Rev. D {\bf 11}, 2461 (1975);
G. Bialkowski, K. Kang and  B. Nicolescu, Lett. Nuovo Cimento {\bf 13}, 401 (1975).
%
\bibitem{bhp}M.~M.~Block, F.~Halzen and G.~Pancheri, e-Print Archive: {\bf hep-ph/0111046 v2}, Eur. Phys. J. C {\bf 23}, 329 (2002).
%
\bibitem{rhogp}M. M. Block, e-Print Archive: {\bf hep-ph/0204048}, Phys. Rev. D {\bf 65},116005 (2002).
%
%
\bibitem{rhogg}M. M. Block and G. Pancheri, e-Print Archive: {\bf hep-ph/0206166}, Eur. Phys. J. C {\bf 25}, 287 (2002).
 
%
\bibitem{bc}  M.~M.~Block and R.~N.~Cahn, Rev.~Mod.~Phys.~{\bf 57}, 563 (1985).
%
%
\end{thebibliography}
\end{document}